\documentclass[10pt,conference]{IEEEtran}
\usepackage{mathtools}
\usepackage{amsmath}
\usepackage{resizegather}
\usepackage{multirow}
\usepackage{balance}
\usepackage{graphicx}
\usepackage{comment}

\newcommand\myfrac[2]{\genfrac{}{}{0pt}{}{#1}{#2}}
\begin{document}
\title{	Quantum Encoded Quantum Evolutionary Algorithm for the \\ Design of Quantum Circuits}

\author{\IEEEauthorblockN{Georgiy Krylov}
\IEEEauthorblockA{Department of Computer Science\\
Nazarbayev University\\
Astana,Kazakhstan\\
Email: gkrylov@nu.edu.kz}\and
\IEEEauthorblockN{Martin Lukac}
\IEEEauthorblockA{Department of Computer Science\\
Nazarbayev University\\
Astana,Kazakhstan\\
Email: martin.lukac@nu.edu.kz}}


%


\maketitle

\begin{abstract}
In this paper we present Quanrum Encoded Quantum Evolutionary Algorithm (QEQEA) and compare its performance against a a classical GPU accelerated Genetic Algorithm (GPUGA). 
The proposed QEQEA differs from existing quantum evolutionary algorithms in several points: representation of candidates circuits is using qubits and qutrits and the proposed evolutionary operators can in theory be implemented on quantum computer provided a classical control exists. The synthesized circuits are obtained by a set of measurements performed on the encoding units of quantum representation. Both algorithms are accelerated in GPGPU. The main target of this paper, is not to propose a completely novel quantum genetic algorithm but to rather experimentally estimate the advantages of certain components of genetic algorithm being encoded and implemented in a quantum compatible manner. The algorithms are compared and evaluated on several reversible and quantum circuits. 
The results demonstrate that on one hand the quantum encoding and quantum implementation compatible implementation provides certain disadvantages with respect to the classical evolutionary computation. On the other hand, encoding certain components in a quantum compatible manner could in theory allow to accelerate the search. This acceleration would in turn counter weight the implementation limitations. 
\end{abstract}


%
\IEEEpeerreviewmaketitle

\section{Introduction}
The direct design of quantum circuits, that is designing quantum or reversible circuits directly using a set of quantum universal gates, suffers from two main problems. First, the optimal method for constructing primitive logic reversible quantum gates for larger number of qubits is not known. The complexity of constructing such gates rises from the fact that component quantum gates applied to at maximum two qubits can be used. Second, designing large reversible and quantum circuits from macros, does not guarantee an exact minimal and optimal design. As a result, there is no real quantum circuit design algorithm for both quantum and reversible circuits using directly quantum primitives. 

The synthesis of reversible quantum gates such as gates from the $C^nU$ family with $U$ being $NOT$, or $SWAP$ unitary operations, has been solved in general for some sets of Turing universal quantum gates and small number of qubits. For instance, the minimal realization of $C^2NOT$ gate is known in the $CNOT/CV/CV^\dagger$, Clifford-T or CH/CZ set of quantum gates. However, in the Ising model the Toffoli gate is not known with certainty as the original specification was found by a stochastic algorithm~\cite{lee06} while in~\cite{7964993} an improved realization was found. Additionally, this situation only gets worse with larger logic gates, where synthesis is done by LUT~\cite{soeken:17} or replacement of large gates by a group of smaller gates already known~\cite{szyprowski:11}. Thus, a synthesis method that designs larger quantum circuits directly using quantum gates would benefit from better minimal cost but also would require faster computers. 

The evolutionary approach is one of the possible way to find cheaper realizations of the $C^nU$ quantum gates. The reason is that while for smaller Turing-universal gate sets and relatively small circuits it is possible to enumerate all possible gate combinations and therefore obtain the less costly minimal gate realization; for larger sets of gates and quantum circuits enumeration would take too long. Consequently, a pseudo-evolutionary search can accelerate the search for more optimal realization of small and medium sized quantum gates using evolutionary operators. 

However, designing algorithm directly for quantum computer is not a trivial task. To do that the specific principles and constraints of quantum computing~\cite{nielsen:00} has to be taken into account. Naturally, one can use classical algorithms implemented and accelerated in quantum computer such as SAT~\cite{dwave} but such algorithms suffer from the overhead of building classical mechanisms using quantum computational elements. Thus, implementing an algorithm that is as close as possible to quantum information and this manipulation using basic quantum tools is most desirable.

In this paper we propose a quantum encoded quantum evolutionary algorithm (QEQEA) parallelized on GPGPU and we compare it to an equivalently GPGPU accelerated classical evolutionary algorithm. QEQEA uses qubits and qutrits to represent parameters evolved by the quantum evolutionary operators as compared to classical genetic algorithm that uses classical strings. The evolutionary operators of QEQEA are strongly simplified and adapted to be quantum-realizable; the used evolutionary operators are built from unitary evolution and measurement process. The QEQEA, evolves simple quantum gates that are used to build the quantum circuits. From one single population of gates, several quantum circuits are sampled by measurement. As such the QEQEA is intended for ensemble quantum computer approach such as NMR~\cite{nielsen:00} or One-Way quantum computing~\cite{raussendorf:03}. Each quantum gate is updated proportionally to fitness values stored in the non-quantum part of the algorithm. The QEQEA and the classical GPUGA are both evaluated on the Ising model of quantum computer~\cite{nielsen:00} due to high complexity and high number of parameters to optimize. The results presented here are aimed to evaluate the difference between these two algorithms rather than provide new state-of-the art circuits realizations in currently used models of quantum gates such as Clifford-T. 



In summary the following quantum-like modifications are implemented in QEQEA:
\begin{itemize}
    \item ensemble-quantum computer inspired set of evolutionary operators,
    \item population of candidates solutions encoded using qubits and qutrits,
    \item adaptive mutation as the main driving force of the evolution,
    \item templates for building interaction gates,
    \item use of position in the memory to encode circuit information
    \item measurement based quantum gate and quantum circuit creation.
\end{itemize}


This paper is organized as follows. Section~\ref{sec:background} introduces required knowledge about genetic algorithms and Section~\ref{sec:qcircuits} introduces the required information about quantum computing and quantum circuits. Section~\ref{sec:model} introduces the proposed model. Section~\ref{sec:results} describe the experimentation and obtained results and Section~\ref{sec:disc} discusses the quantum implementation discrepancies and performance considerations of the proposed algorithm. Finally Section~\ref{sec:conclusion} concludes the paper. 
\section{Background}
\label{sec:background}
The genetic algorithms are a class of pseudo-random search algorithms inspired by natural evolution. They are used to solve difficult search and optimization problems that 
are otherwise unsolvable by exhaustive or analytic approaches in reasonable amount of time or with bounded memory. 
The application of evolutionary algorithm for quantum synthesis is a topic that has been previously studied since the beginning of the century \cite{1029883,dill:98,hadjam:10}.  A lot of work has been done for solving the problem in classical paradigm using different approaches and hardware, however the execution time is a limiting factor even for the most optimal evolutionary and general algorithms~\cite{lukac:12} directly designing quantum circuits. Thus, Quantum and Quantum Inspired Algorithms were introduced in order to reduce the computation time using principles of quantum mechanics. One of the first evolutionary algorithm inspired by quantum computing was developed in~\cite{quantum-inspired-first}. The most original idea was the extension of quantum inference crossover~\cite{quantum-inspired-first}. In~\cite{quantum-inspired-GA} the first definition and requirements for evolutionary quantum algorithms have been introduced. 
The most important and challenging requirements are listed below for the clarity of understanding:
\begin{itemize}
    \item A reasonable method of splitting the problem to sub-problems
    \item "The number of universes required should be identified" \cite{quantum-inspired-GA}, that is the number of quantum registers should be well described
    \item The computations should occur in parallel
    \item "There must be some form of interaction between all of the universes. The interference must either yield a solution, or new information for the universes to utilize in locating a solution" \cite{quantum-inspired-GA}
\end{itemize}
Several further studies described the Quantum Genetic Algorithms for general purposes \cite{6236542} \cite{4358783} such as for the knapsack problem. The problem of quantum circuits synthesis was studied using Quantum Evolutionary Algorithm (QEA) in \cite{Ding2008}. The study\cite{Ding2008} used integer representation of population, and demonstrated synthesis with multiple controlled $NOT$ gates. In~\cite{1107.3383} the design of quantum circuits used qutrits for individual encoding. This allowed for more advantageous usage of mutation and ternary operators. 
In order to run these algorithms, most of the studies design special quantum encoding and mapping of evolutionary operators that could potentially allow to execute their algorithm on quantum computers. 
The simulation task is not trivial and requires optimization and performance acceleration by itself. Building quantum simulators is also developing because of the need of benchmark and corrections for upcoming quantum hardware  \cite{1704.01127}.

The previous studies in the quantum and in quantum inspired evolutionary computation fields outline several possible improvements. For problems dealing with higher dimensional space such as multi-qubit complex vector space (Hilbert space) operators efficiency, the selection methodology and fitness evaluation should be evaluated for both performance and accuracy w.r.t to its classical counter parts. 


\section{Quantum Circuits and Quantum Gates}
\label{sec:qcircuits}

Information in quantum circuit is represented by a qubit $\vert\phi\rangle$ represented by a wave state $\vert\phi\rangle = \alpha\vert0\rangle+\beta\vert1\rangle$. Multiple qubits are expanded into a quantum register using Kronecker product such as for two qubits $\vert a\rangle$ and $\vert b\rangle$ the joint state is 
\begin{equation}
\vert\psi\rangle = \vert a\rangle\otimes\vert b\rangle = \alpha_a\alpha_b\vert00\rangle+\alpha_a\beta_b\vert01\rangle+ \beta_a\alpha_b\vert 10\rangle+\beta_a\beta_b\vert11\rangle
\end{equation}
The logic operations applied upon qubits are specified by unitary matrices. For instance, to negate the qubit $\vert a\rangle$ the $X$ operator can be applied (note the coefficients reordered): 
\begin{equation}
    X\vert\psi\rangle = \beta_a\alpha_b\vert00\rangle+\beta_a\beta_b\vert01\rangle+ \alpha_a\alpha_b\vert 10\rangle+\alpha_a\beta_b\vert11\rangle 
\end{equation}

\begin{figure}[bht]
\centering
\includegraphics[width=0.5\linewidth]{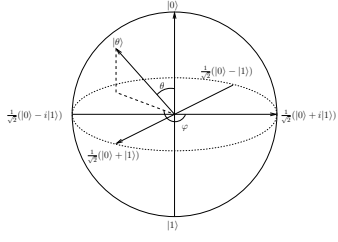}
\caption{\label{fig:bloch} Bloch sphere}
\end{figure}

The Ising model of quantum computing~\cite{nielsen:00} uses three single qubit quantum gates $R_X(\theta)$, $R_Y(\theta)$ and $R_Z(\theta)$ and one two-qubit interaction quantum gate $I_{ZZ}(\theta)$. The gates are parameterized by an angle of rotation $\theta$ from the range $[0,2\pi]$. The rotations applied to a single qubit can be visualized using the unitary Bloch sphere on Figure\ref{fig:bloch}

\begin{figure}[bht]
\centering
\includegraphics[width=0.8\linewidth]{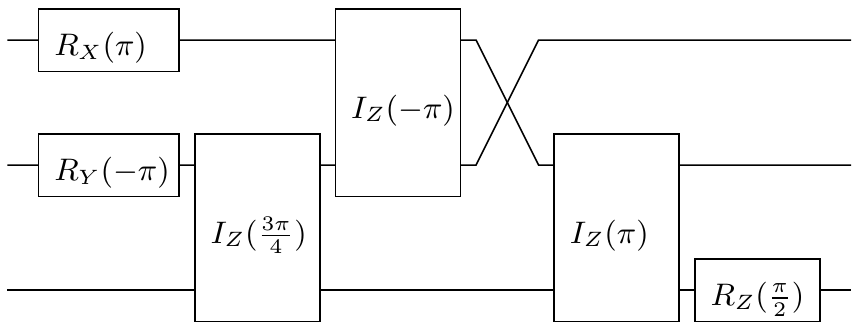}
\caption{\label{fig:qcirc} Example of arbitrary quantum circuit in the Ising model}
\end{figure}

A sequence of single and two-qubit operators (gates) applied to a quantum register is called a quantum circuit. Example of a quantum circuit is shown in Figure~\ref{fig:qcirc}.

%
The single qubit rotations can be described by following equations\cite{lee06}:
  \begin{itemize}
	  \item X direction
  \begin{gather}
  R_x(\theta)=e^{\left (\frac{-i\theta X}{2} \right )}=cos\left(\frac{\theta}{2}\right)I_2-i sin\left(\frac{\theta}{2} \right)X =\begin{bmatrix} cos\left(\frac{\theta}{2}\right)&-i sin \left(\frac{\theta}{2}\right)\\-i sin \left(\frac{\theta}{2}\right)& cos \left(\frac{\theta}{2}\right)\end{bmatrix}
  \end{gather}

	  \item   Y direction:
  \begin{gather}
  R_y(\theta)=e^{\left (\frac{-i\theta Y}{2} \right )}=cos\left(\frac{\theta}{2}\right)I_2-i sin\left(\frac{\theta}{2}\right) Y =\begin{bmatrix} cos\left(\frac{\theta}{2}\right)&- sin \left(\frac{\theta}{2}\right)\\ sin \left(\frac{\theta}{2}\right)& cos \left(\frac{\theta}{2}\right)\end{bmatrix}
  \end{gather}

	\item   Z direction:
  \begin{gather}
	  R_z(\theta)=e^{\left (\frac{-i\theta Z}{2} \right )}=cos\left(\frac{\theta}{2}\right)I_2-i sin\left(\frac{\theta}{2}\right) Z =\begin{bmatrix} e^{-i\left(\theta/2\right)}&0\\ 0& e^{i\left(\theta/2\right)}\end{bmatrix}
  \end{gather}

  \end{itemize}

  The template for the two-qubit interaction in Ising model \cite{lee06}:
  \begin{equation}
	  J_{ij}(\theta)=e^{\frac{-i\theta}{2}} \begin{bmatrix}1&0&0&0\\0& e^{\theta}&0&0\\ 0& 0&e^{\theta}&0\\0&0&0&1\end{bmatrix}
  \end{equation}


\section{Quantum Evolutionary Algorithm description}
\label{sec:model}

The proposed approach is distinguished from previous work in the following points:
\begin{itemize}
    \item synthesis on a level of single qubit rotations and two-qubit interaction gates,
    \item unique encoding of quantum population using qubits and qutrits,
    \item GPU accelerated version for simulation of quantum computer behavior; we offer unique mapping of quantum operators that enables optimization for GPU acceleration,
    \item predefined templates of interaction matrices for simplification of the search,
    \item evolutionary operators are a combination of adaptive mutation and SU(3) rotations (in case of qutrits).
\end{itemize}

\begin{figure}[bht]
\centering
\includegraphics[width=0.6\linewidth]{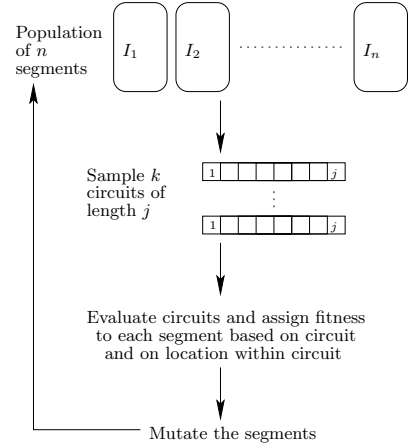}
\caption{\label{fig:flow} High level flow of the quantum evolutionary algorithm}
\end{figure}

The proposed algorithm is briefly depicted in Figure~\ref{fig:flow}. The QEQEA does not evolve circuits directly; instead a set of quantum gates (segments) are evolved as a population. The circuits are obtained by probabilistic selection of gates from the population. Each gate is encoded by several quantum parameters and uses measurement procedure for circuit construction. 

\subsection{Quantum Gates Representation}

In this work the Ising model of quantum circuits is used. While this model is only theoretically useful, it is the most complex. Generating results in this circuit model is an indicator of the performance of the applied algorithm.  There are two types of primitive gates in Ising model: rotations and interactions that were described in Section~\ref{sec:qcircuits}, and they were used in the QEQEA.
In the QEQEA, each quantum gate is represented by a set of qubits and /or qutrits.

\subsubsection{Rotation gates}
The single qubit gates ($R_X(\theta)$, $R_Y(\theta)$ and $R_Z(\theta)$) are encoded using one qubit and one qutrit. The angle of rotation $\theta$ is represented by the qubit parameter specifying its complex amplitudes: $e^{-i\pi\theta}$ . The axis of rotation is obtained by measuring the state of the qutrit. We repeat the measurement process multiple times to approximate the state of the qutrit, without eliminating uncertainty. The qutrit states:
$\{\vert0\rangle, \vert1\rangle, \vert2\rangle \}$ correspond to rotations around $\{x,y,z\}$ axis, respectively.
\subsubsection{Interaction gates}
The second type of quantum gate we use is the two-qubit interaction. The interaction gate is equivalent to two parameterized $Z$ rotation gates applied simultaneously to two qubits~\cite{lee06}. By introducing interaction matrices templates, we reduced the number of parameters required to construct the interaction gate to one. This parameter is the angle of rotation is obtained by copying qubit value similar to the case of single qubit gates construction. For $numberOfWires$ wires, there can be at most $\left (\myfrac{numberOfWires}{2}\right )$ possible configurations (we call them templates).

\subsubsection{Interaction Gate Templates}

The interaction gate can be expressed in form of term-wise exponent of scalar multiple of gate parameter value and special diagonal matrix. The elements of this diagonal matrix are $+1$ and $-1$ depending on the wires on which the interaction is applied. 
Notice that the number of possible interaction templates grows slowly considering the size of the problem. Thus, all possible templates can be  stored in memory. When an interaction gate is to be inserted in a quantum circuit, parallel exponent of elements of diagonal matrices instead of matrix multiplications between swap gates and interactions are performed. 
\begin{figure*}[thb]
\includegraphics[width=\linewidth]{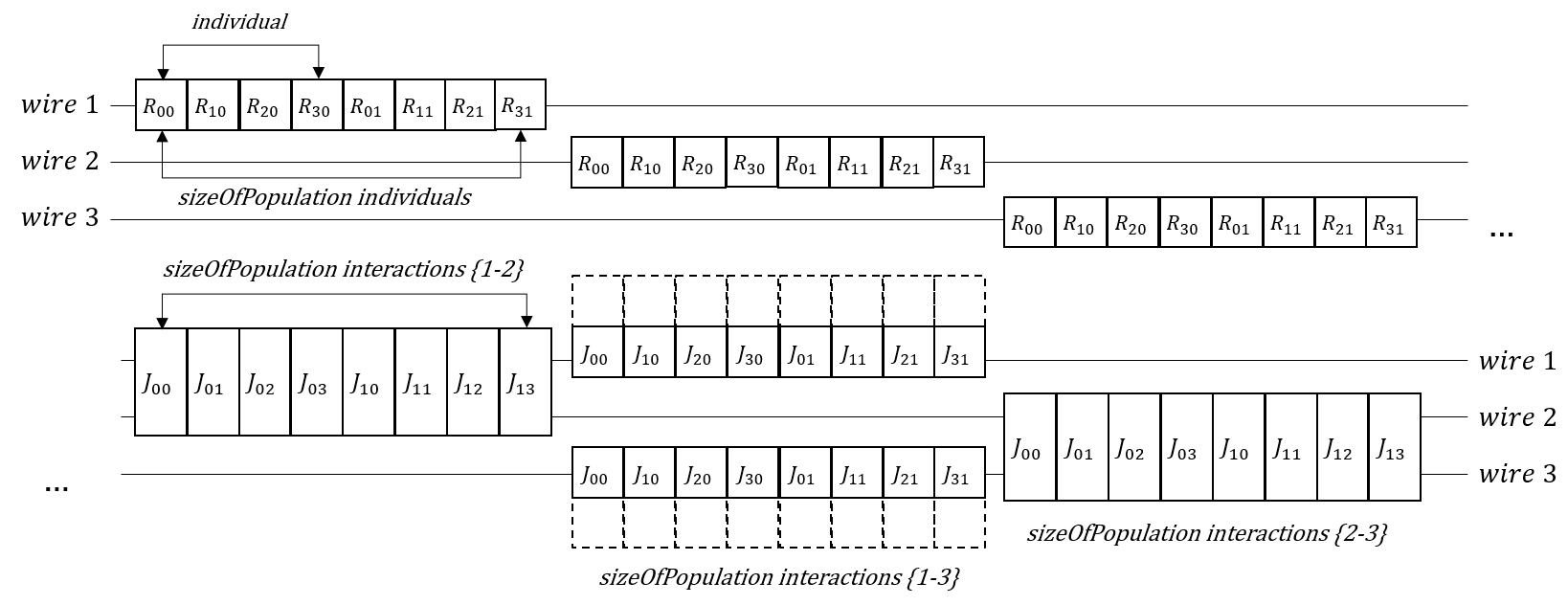}
\caption{\label{fig:popExample} Example of memory allocation for three-qubit circuit of length four with two individuals in the population}
\end{figure*}
These templates allow to exclude swap gates 
for interaction between non-neighboring qubits and therefore gives potential to synthesize more optimal circuits.



\subsection{Population Initialization}
\label{sec:popinit}
In the proposed QEQEA, the evolution is performed by local modification of a set of qubits and qutrits. These elementary quantum information units encode a set of quantum gates that are sampled into several possible quantum circuits. 
The set of evolved qubits (representing the population) can be conditionally split to two regions : single qubit rotations and interactions region. The population of the qubits and qutrits is defined by following parameters
\begin{itemize}
    \item $sizeOfIndividual$ – parameter corresponding to the length of the circuit in gates (segments)
    \item $sizeOfPopulation$ – parameter corresponding to number of individuals(circuits) in the evolution. The segments count in population grows to:
    \begin{equation}
        sizeOfPopulation*sizeOfIndividual
    \end{equation} 
    \item $numberOfWires$ - parameter corresponding to the number of qubits of the target quantum gate. From this parameter, the $interactionTemplatesNumber$ is derived. Thus,$numberOfWires$ parameter increases the amount of memory required to store the qubits population to:
    \begin{multline}
     (interactiontemplatesChose+numberOfWires)\\ *sizeOfPopulation*sizeOfIndividual
     \end{multline}
   
\end{itemize}
The number of qutrits is fixed at
\begin{multline}
numberOfWires * sizeOfPopulation * sizeOfIndividual,
\end{multline}because qutrits are used only for segments that represent single qubit rotations.
 
The number of individuals in the population raises the amount of initial information to explore. It also significantly increases the computational complexity. However, the tasks required to synthesize one individual may be executed in parallel and we aim to get most benefit of highly effective parallel capabilities of quantum computers at that stage of the algorithm. 


Figure~\ref{fig:popExample} describes an example population that would have two individuals, targeting to synthesize the circuit consisting of four gates applied to three input qubits (wires). The first eight qubits encoding the circuit segments correspond to rotation applied on the first input wire (labeled "wire 1" in Figure~\ref{fig:popExample}). There are exactly eight qubits in this particular case because the population consists of two individuals of size four. Similarly, the next eight qubits correspond to rotation on the second wire (labeled "wire 2" in Figure~\ref{fig:popExample}). Same rules apply to the third set of eight qubits. The remaining twenty four qubits do not have qutrits allocated for them because they belong to interaction region and use pre-calculated templates instead of measured axis (labeled "Interactions" in Figure~\ref{fig:popExample}). The figure does not contain qutrits in it, the qutrits are described on the Figure~\ref{fig:SC2}. 

\subsection{Segments construction}
\label{sec:segmConst}
The population of qubits and qutrits encodes a set of segments. Each single qubit rotation gate is expanded to the width of the full circuit  defined by $numberOfWires$ parameter. For interaction gates, templates are expanded to the circuit width before the evolutionary process starts and then are simply retrieved from memory. The obtained segments (rotation and interaction) are used to build the target candidate circuit. The restriction of using only one quantum gate per segment allowed for more efficient acceleration on the GPGPU~\cite{7964993}. 
Additionally, as described in Section~\ref{sec:popinit}, the population uses structure to encode following properties 
\begin{itemize}
    \item 
    The wire on which the gate should operate
    \item
    The position of a segment inside of the individual
\end{itemize}
\begin{figure}[bht]
\includegraphics[width=\linewidth]{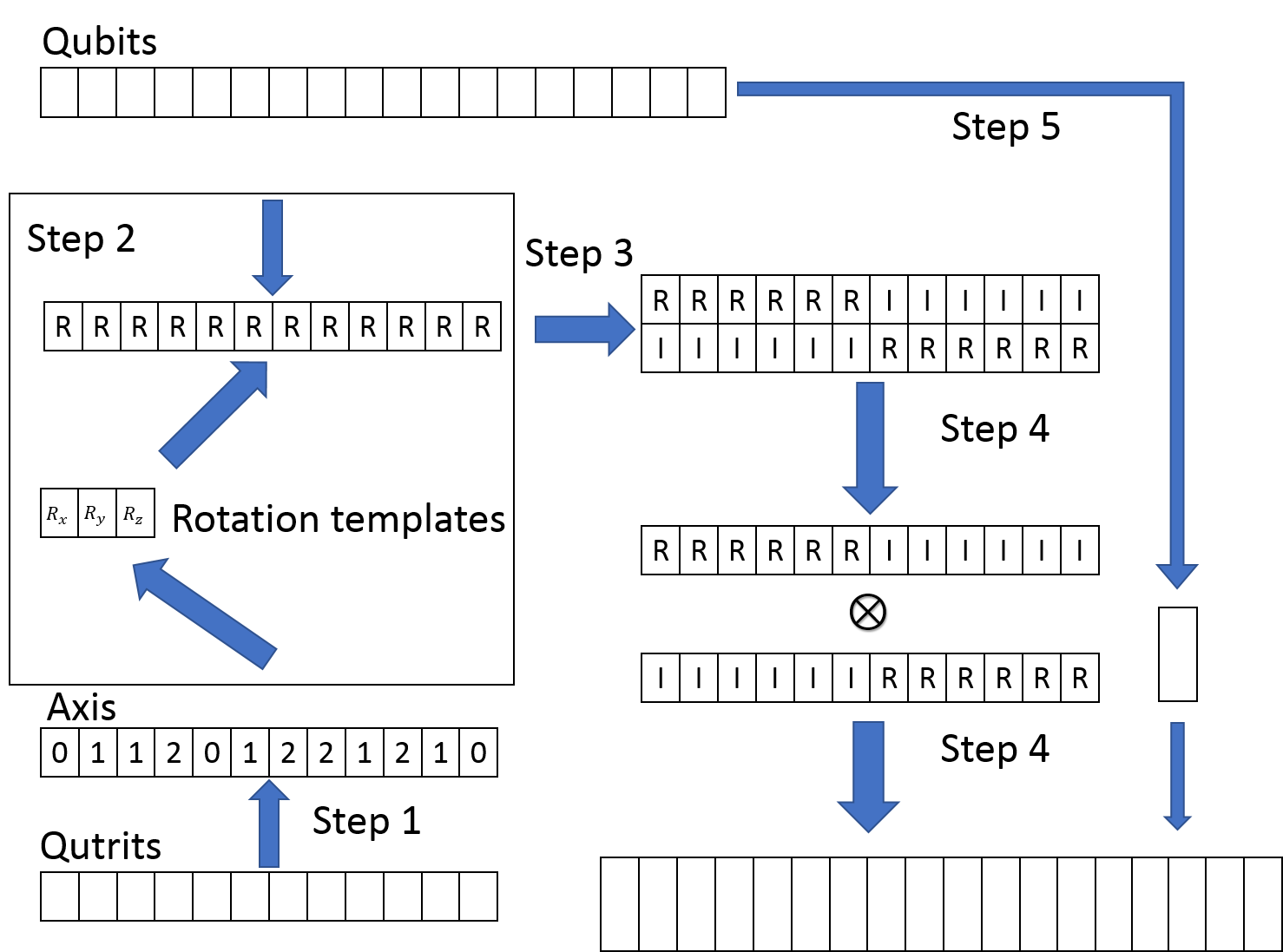}
\caption{\label{fig:SC2} Process of Segment Construction}
\end{figure}
The segment construction procedure consists of five steps and is illustrated in Figure~\ref{fig:SC2}.
\begin{itemize}
    \item Step 1: Measure the qutrits to get an array of axes of rotation
    \item Step 2: Plug in the qubits values for corresponding matrix template determined by qutrit measure and parameter value defined by qubit
    \item Step 3: Rearrange the memory in parallel Kronecker Product friendly order
    \item Step 4: Apply Kronecker product simultaneously to rotation matrices
    \item Step 5: Construct interaction matrices in a way it was described above and put next to segments obtained from rotation matrices
\end{itemize}

Note that the allocated memory represents directly units of quantum information. Thus, references to memory allocation represent directly qubit and qutrit allocation in particular order. 




%


\subsection{Circuit construction}
\label{sec:segs}

The proposed quantum circuit design method is a form of evolutionary algorithm heavily altered in order to allow some of its components to be directly mapped into a quantum computer. Additionally, the proposed algorithm is also intended to be efficiently implementable on a highly parallel device such as GPGPU. 



\begin{figure}[bht]
\centering
\includegraphics[width=0.7\linewidth]{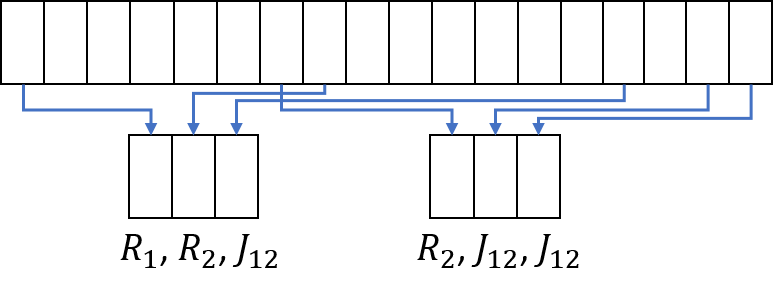}
\caption{\label{fig:CC} Building a circuit of length three affecting two qubits having two individuals in the population}
\end{figure}

For the circuit of length of $sizeOfIndividual$ we launch $sizeOfIndividual$ parallel threads each indexed by $index_{thread}$. Each thread generates two random numbers: 
\begin{itemize}
    \item 
    $whichIndividual$ from range $0..sizeOfPopulation$
    \item
    $whichRotationOrInteraction$ from the range \newline$0..numberOfWires+interactionTemplatesNumber$
\end{itemize} These values are later used to calculate the index of segment to be plugged in the circuit:
\begin{multline}
    segmentIndex=whichRotationOrInteraction\\ *sizeOfIndividual*sizeOfPopulation\\ +whichIndividual*sizeOfIndividual+index_{thread}
\end{multline}
The result of calculation is stored as reference to a segment in population for $index_{thread}$ position in the circuit. An example of process is shown in the Figure~\ref{fig:CC}. 


We repeat this process $sizeOfPopulation$ times to generate $sizeOfPopulation$ circuits each iteration.

\subsection{Fitness Evaluation}
  The fitness value reflects the proximity of the synthesized circuit matrix $S$, to the target circuit matrix $T$. The possible values of selected function are ranging from 0 to 1, where 1 represents identical matrices and 0 being the opposite. The approach is based on property of unitary matrices that $U^\dagger*U = I$. The Equation \ref{eq:distanceMeasure} displays the actual fitness function:
\begin{equation}
\label{eq:distanceMeasure}
fitnessValue(S)=1-\sqrt{\frac{size-|tr(S^\dagger T)|}{size}}
\end{equation}
In this expression, the $||$ operator corresponds to modulo operation. The $tr$ operation represents calculation of the sum of diagonal elements. The $size$ is normalization constant and is taken to be equal to $2^{numberOfWires}$.
\subsubsection{Segment Fitness}

 Each segment used during circuit construction stage (Section~\ref{sec:segs}) is assigned with a fitness value. 
 The fitness value assigned to each segment is the same as the fitness value of the circuit it was used to construct. 
 
 Additionally, an elitist approach was implemented: if the new fitness value of a segment is better than the previous best value, the states of the qubits and qutrits are preserved, otherwise they get discarded. 
 
 Finally, each segments fitness is tied to a particular position in a given circuit. That is, the same segment will be represented by various fitness values depending on the position where it was located within the synthesized circuit.

\subsection{Evolutionary Search}


The main differences between the classical GPUGA and the proposed QEQEA is the mutation operation and the lack of crossover. We use adaptive mutation inspired from the evolutionary strategies approach~\cite{AUGER200535}. 
The mutation is proportional to the error, i.e. better individuals undergo less significant changes\cite{Hong:2014:SSB:2598394.2609873}. This approach is argued to be more effective than the mutation with constant probability and mutation range\cite{MarsiliLibelli2000}. Additionally, the $probabilityOfMutation$ parameter is introduced in the algorithm to make the mutation operation probabilistic. 

Each individual undergoes change per iteration of our algorithm with $probabilityOfMutation$. Every time the mutation is to be performed, there are two equiprobable operations that may happen: qubits or qutrits mutation. 

\begin{itemize}
\item 
We use the $mutationRange$ parameter that determines the maximum possible change to qubit parameter. In our algorithm, it is taken to be fraction of $\pi$. The formula to calculate the mutation value is $\pm(1-segmentFitness)*mutationRange$. The qubit parameters are assumed to stay within $[0,2*\pi]$ range, so after the mutation the resulting parameter is readjusted modulo $2*\pi$
\item The qutrits mutation is performed by applying the arbitrary SU(3) rotations on a qutrit \cite{PhysRevA.85.032331}. Such matrix can be generated using eight parameters: three rotation angles $\theta_1,\theta_2,\theta_3$ from range $0<\theta<\pi/2$ and five phases $\phi_1,\phi_2,\phi_3,\phi_4,\phi_5$ from range $0<\phi<2*\pi$. Equation~\ref{eq:su3} shows the template used to calculate the mutation on the qutrits, with $c_{k}=cos\theta_{k}$ and $s_{k}=sin\theta_{k}$.
\end{itemize}
\begin{figure*}
\begin{equation}
\textbf{U} = \begin{bmatrix}e^{i\phi_1}c_1c_2 & e^{i\phi_3}s_1 &  e^{i\phi_4}c_1s_2 \\ 
e^{-i\phi_4-i\phi_5}s_2s_3 - e^{i\phi_1+i\phi_2-i\phi_3}s_1c_2c_3 &
e^{i\phi_2}c_1c_3 &
-e^{-i\phi_1-i\phi_5}c_2s_3 - e^{i\phi_2-i\phi_3+i\phi_4}s_1s_2c_3 \\
-e^{-i\phi_2-i\phi_4}s_2c_3 - e^{i\phi_1-i\phi_3+i\phi_5}s_1c_2s_3 &
e^{i\phi_5}c_1s_3  &
e^{-i\phi_1-i\phi_2}c_2c_3 - e^{-i\phi_3+i\phi_4+i\phi_5}s_1s_2s_3 
\end{bmatrix}
\label{eq:su3}
\end{equation}
\end{figure*}

During one step of mutation, one of these nine parameters is generated randomly from a domain of its possible values multiplied by $1-segmentFitness$. The constructed operator is then applied to the target qutrit. 

The crossover operation was intentionally removed from the model, as our genotype - the array of qubits and qutrits is used to generate a population of circuits. Thus, only one set of qubits and qutrits is evolved and the crossover is replaced by the location dependent segment fitness value.

    
\section{Results}
\label{sec:results}
\subsection{Evaluation of QEQEA}
To verify the QEQEA algorithm we tested it on several small quantum gates: $C^2NOT$, Peres and $CNOT$. Table~\ref{tab:1} shows the results of the search for the $CNOT$ gate. 

The Table \ref{tab:1} presents the outputs from the algorithm obtained in the process of synthesizing a CNOT gate. Each row in the table from top to the bottom represent encoded circuit segments in the order they appear in the synthesized circuit. Each row of the table contains all information required to decode information about circuit segment. The first column contains the parameter value $\theta$ representing the rotation. The second column determines whether the parameter $\theta$ should be plugged to rotation or interaction template.
The third column of the table contains the states of the qutrit, which after measurement indicate the direction of the rotation gate. The value of this column should be ignored if the segment is a two-qubit interaction. 
The fourth column indicates the axis of rotation obtained as a result of measurement.


\begin{table}[bht]
\centering
\caption{\label{tab:1}Result of CNOT gate synthesis ($sizeOfIndividual$=3, $sizeOfPopulation=1$)}
\begin{tabular}{c c c c }
\hline
Parameter $\theta$  & Index in memory & Qutrit states & Axis\\ 
 \hline\hline
$\pi/2$ & 0 &  -0.43 - 0.16$i$;  0.85 + 0.08$i$; 0.03 - 0.24$i$&y\\ 
$3\pi/2$ & 7 & Interaction template between 1 and 2 & \\  
$3\pi/2$& 2 &  0.39 - 0.66$i$; -0.43 + 0.43$i$; 0.16 - 0.14$i$ & x   
\end{tabular}
\end{table}
thus,Table~\ref{tab:1} represents a CNOT circuit constructed using the following sequence of gates:$R_{1y}(\theta=1.570796)J_{12}(\theta=4.712389)R_{1x}(\theta=4.712389)$. 

\begin{figure}[bht]
\centering\includegraphics[width=0.4\linewidth]{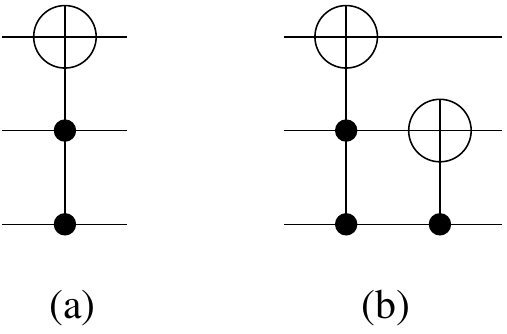}
\caption{\label{fig:gates1} a) Representation of Toffoli and b) Peres logic gates}
\end{figure}

Table~\ref{tab:2} shows the resulting matrix of the obtained $C^2NOT$ gate with the length of 16 segments. The schematic representation of the target Toffoli gate is shown in Figure~\ref{fig:gates1}a). Some terms of the matrix have differences from original Toffoli gate therefore the circuit obtained is not exact, however on average the error per term is $\approx ~0.02$. The reason of not having exact gate appeared due to the convergence to local maximum.

\begin{table}[bht]
\centering
\caption{\label{tab:2}Result of Toffoli gate synthesis ($sizeOfIndividual$=16)}
\begin{tabular}{ c c c c c c c c }
0.894 & 0.000 & 0.004 & 0.000 & 0.101 & 0.000 & 0.000 & 0.000\\
0.000 & 0.916 & 0.000 & 0.001 & 0.000 & 0.080 & 0.000 & 0.004\\
0.004 & 0.000 & 0.967 & 0.000 & 0.000 & 0.000 & 0.029 & 0.000\\
0.000 & 0.004 & 0.000 & 0.121 & 0.000 & 0.000 & 0.000 & 0.875\\
0.101 & 0.000 & 0.000 & 0.000 & 0.894 & 0.000 & 0.004 & 0.000\\
0.000 & 0.080 & 0.000 & 0.004 & 0.000 & 0.916 & 0.000 & 0.001\\
0.000 & 0.000 & 0.029 & 0.000 & 0.004 & 0.000 & 0.967 & 0.000\\
0.000 & 0.000 & 0.000 & 0.875 & 0.000 & 0.004 & 0.000 & 0.121\\
\end{tabular}
\end{table}


\subsection{Comparing QEQEA and GA}

The comparison of the performance was done between the QEQEA and the non-quantum GPUGA algorithm from~\cite{7964993}. The reason to compare QEQEA with algorithm from ~\cite{7964993} is that the GPUGA provides similar algorithmic and acceleration basis for comparison. In fact, the QEQEA was developed as a quantum extension of the original non quantum algorithm. The main differences are:
\begin{itemize}
    \item Representation: same mapping from memory to individual was implemented. The representation of quantum gate (segment) was performed using a set of real and complex coefficients
    \item The Evolutionary operators: two point crossover was used and the mutation was a random small alterations of the gate parameters. 
    \item Selection was using the Stochastic Universal Sampling (SUS)
    \item Evolution occurred on the level of level of circuits, not on the individual gates (segments). 
    \item In the GPUGA no qutrits were used; we introduced the qutrits in QEQEA in order to avoid allocating extra memory for each type of the rotation gates (x,y,z) direction. This evolution of qutrits could possible reduce computation time required for each population step. 
\end{itemize}

The common parts of both algorithms are in the GPU acceleration and parallelism. The computational overhead that was required for the implementation of the QEQEA is the amount of measurement used during the creation of candidate segments from the encoding qubits and qutrits. Despite these various implementation differences the two algorithms are evaluated for speed of convergence and ability to find the desired solution as many of their components were programmed in a similar manner.

The Table~\ref{tab:compare1} shows the differences of speed in obtaining the various gates for which we tested both algorithms. Notice that in all cases the classical algorithm was faster than the QEQEA algorithm (iteration of QEQEA takes significantly more time). Thus even if the iteration number is smaller in QEQEA, the GPUGA is faster in real time and was able to converge to better results. The reason is the fact that the QEQEA is evolving gates rather than whole circuits while the classical GA evolves whole circuits. Additionally, the QEQEA generates solutions from a single set of encoding qubits and qutrits. As such there is no crossover because there is only one  individual of qubits and qutrits. Consequently, because the main evolution mechanisms are selection and mutation, the proposed QEQEA is more related to evolutionary strategies rather than to genetic algorithm.  

\begin{table}[bht]
\centering
\caption{\label{tab:compare1} Comparison of Results and performance between the QEQEA and a classical GPGPU}
\begin{tabular}{|c|c|c|c|c|}
\hline
\multirow{2}{*}{Function}&\multicolumn{2}{c|}{QEQEA}&\multicolumn{2}{c|}{GPGPU}\\
\cline{2-5}
& Accuracy& No. Generations& Accuracy& No. Generations\\
\hline
\hline
CNOT & 1.0000  & 400 & 1.0000 &200\\
Toffoli& 0.7047 & 13000  & 0.9663 & 34500\\
CCCNOT& 0.6464 & limit & 0.7539 & 650970\\
Peres &0.5693 & limit & 0.9443 & 2M\\
\hline
\end{tabular}
\end{table}
The first and the third columns of Table \ref{tab:compare1} display the accuracy of best results achieved by each algorithm. The iterations number could also serve as a measure for performance comparison, however for the QEQEA this data is only partially available. The reason for that is the search of CCCNOT and Peres gates reached the maximum iterations limit of ten million iterations. However, this fact also means the result could be possibly improved if the higher limit for iterations was set. Careful reader may notice the difference between best available results shown in this table and in \cite{7964993}. This is due to re-evaluation of accuracy of previously achieved results with respect to the new fitness function described above. 

\section{Discussion on the performance and realism of the implementation}
\label{sec:disc}
While the search for the gates was partially successful (and thus, confirming that the proposed approach converges), the main drawback of the QEQEA was the slower convergence due to the simulated quantum evolution of individual gates. However, the changes implemented are intended to simulate the implementation of the QEQEA using certain quantum components and thus, the main concern was the general convergence and feasibility.  

In more details the following design choices of implementation of the QEQEA affected its overall performance. 

\begin{enumerate}
    \item First, the structure of the task requires our population being represented as floating point numbers. While this is a limitation for GPGPU it is an advantage for quantum computer where each qubit and qutrit exists in a state-wave state. Using quantum computer an amplitude estimation technique would have to be used in order to estimate the parameter $\theta$. 
    \item Second, the use of adaptive mutation approach. The lack of crossover benefits from GPU acceleration, but the algorithm shows performance reduction compared with regular GPU accelerated genetic algorithms~\cite{7964993}. 
    \item Third, the genetic algorithms utilizing elitism approach are more in danger of convergence to local maxima compared with the other approaches, but the specificity of the task enforces us to use this approach to preserve control over the population. 
    \item Additionally, observe that in general the QEQEA is less performant compared to the GPGPU (Table~\ref{tab:compare1}). The main reason is due to the fact QEQEA evolves quantum gates and thus, each gate have fitness representing it general usefulness rather than its usefulness in a particular quantum circuit. 
\end{enumerate}

Several design choices making QEQEA computationally tractable prevent it from being directly ported to a quantum computer. 
The two most important restrictions imposed that would require several changes to the algorithm in order to make it quantum implementable are:
\begin{enumerate}
    \item The measurement approach: for each circuit the qubits are measured and new quantum gates are generated while the unmeasured states of the qubits and qutrits are evolved. In a more realistic setting of the proposed method is to use weak measurement~\cite{oreshkov:05} that would allow to preserve the quantum states at least partially. While the usage of protective measurement directly on the encoding qubits and qutrits makes it impossible for the algorithm to be considered implementable on quantum computer (one would need an infinite amount of copies and evolve them in parallel) it allowed us to at least simulate the result of such process. 
    \item The evolution of quantum gates instead of the whole circuits. This was decided in order to avoid entanglement between elements of the quantum circuits in space and in time. While entanglement could be highly beneficial to quantum evolution it also makes the simulation of the evolution much more complex and computationally expensive. 
\end{enumerate}

\section{Conclusion}
\label{sec:conclusion}
We introduced a QEQEA as a means for the synthesis of quantum circuits and we compared its performance with a parallelized version of classical GA. The QEQEA features certain components being a possible target for implementation in a quantum computer but in order to keep the implementation computationally tractable several design choices that made it impossible to port directly to a quantum computer. 

Additionally even if all components of the algorithm were made quantum-implementation compatible, many components would remain classic. In particular this means, that even if QEQEA evolutionary components are mapped to a quantum computer, fitness function values, circuit information, algorithm flow control and other parameters require to be kept in a classical memory. 

The comparison with the classical GPUGA showed that the quantum evolutionary model shows worse performance than the classical evolution. The inferior performance is due to many constraints included in the QEQEA that resulted in strong simplification of the evolutionary process. Consequently the main result is that the evolutionary process for computation as originally proposed in~\cite{holland:92} seems to be most efficient when implemented in classical computer. In quantum computer, an efficient implementation requires the entanglement that would made the search much more efficient. However simulating such system on classical computer requires high computational resources and is not easily compared to a classical GA.

As future work, we plan to integrate weak-measurement for circuit generation and an additional mechanism in classical computer that keeps track of gates fitness with respect to a all circuits it was used to build. An even further improvement is to use more complex encoding such as qudits with higher number of bases states and evolve whole quantum circuits.



\bibliographystyle{IEEEtran}
\balance
\bibliography{main}
\end{document}